# Wind-induced drift of objects at sea: the leeway field method[1]

Øyvind Breivik*
Norwegian Meteorological Institute, Norway

Arthur A Allen
US Coast Guard, Office of Search and Rescue, CT, USA

Christophe Maisondieu
IFREMER, France

Jens Christian Roth
Royal Norwegian Navy, Norway

*Corresponding author address:* Øyvind Breivik, The Norwegian Meteorological Institute, Alleg 70, NO-5007 Bergen, Norway, oyvind.breivik@met.no

Abstract

A method for conducting leeway field experiments to establish the drift properties of small objects (0.1-25 m) is described. The objective is to define a standardized and unambiguous procedure for condensing the drift properties down to a set of coefficients that may be incorporated into existing stochastic trajectory forecast models for drifting objects of concern to search and rescue operations and other activities involving vessels lost at sea such as containers with hazardous material.

An operational definition of the slip or wind and wave-induced motion of a drifting object relative to the ambient current is proposed. This definition taken together with a strict adherence to 10 m wind speed allows us to refer unambiguously to the *leeway* of a drifting object. We recommend that all objects if possible be studied using what we term the *direct method*, where the object's leeway is studied directly using an attached current meter.

We divide drifting objects into four categories, depending on their size. For the smaller objects (less than 0.5 m), an indirect method of measuring the object's motion relative to the ambient current must be used. For larger objects, direct measurement of the motion through the near-surface water masses is strongly recommended. Larger objects are categorized according to the ability to attach current meters and wind monitoring systems to them.

The leeway field method proposed here is illustrated with results from field work where three objects were studied in their distress configuration; a 1:3.3 sized model of a 40-ft Shipping container, a World War II mine and a 220 l (55-gallon) oil drum.

# 1. Introduction

Drifting objects go missing for a variety of reasons, and localizing and recovering them is a central part of search and rescue and pollution mitigation efforts. Recently, several operational forecast models for predicting the evolution of search areas for search and rescue (SAR) objects and objects containing hazardous material (HAZMAT) have appeared (Hackett *et al*, 2006; Eide *et al*, 2007; Breivik and Allen, 2008; Davidson *et al,* 2009). These models rely on Monte Carlo (stochastic) techniques to compute probabilistic search areas from ensemble trajectory models that take into account near-surface currents and the wind field as well as object-specific drift properties to advect the drifting object.

The oceanographic and meteorological forcing fields for the trajectory models may differ substantially. Surface current estimates can come from tidal constituents or a handful of observations or from baroclinic high-resolution ocean models. Likewise, the wind forcing may range from measurements taken at a local meteorological station to detailed wind fields from a numerical weather prediction model where coastal features are well resolved. As small objects respond almost linearly to changes in the wind speed (Breivik and Allen, 2008), the trajectory models usually assume a linear relation between 10-m wind speed and the object's motion through the water. In general, an object's motion through the ambient water masses (referred to as its slip, windage or leeway) is roughly inversely proportional to its immersion ratio, as has been shown by Geyer (1989) and O'Donnell *et* al (1997) and exemplified for the case of shipping containers by Daniel *et al* (2002). This means that an estimate of the leeway speed of an object can be derived from the immersion ratio. Unfortunately, an object on the sea surface with a more complex geometry than simple radial symmetry (a sailboat with no way on being an extreme example) will exhibit substantial crosswind motion. The relation between this crosswind motion and the immersion ratio is not straightforward and will vary greatly from one object to another. As both the downwind and crosswind components of the slip or leeway are needed to compute the trajectory it is crucial to the forecasting of SAR and HAZMAT objects to have reliable estimates of both. This explains why simple linear regression coefficients (discussed in Section 2) have proved the preferred method of parameterizing the leeway of distressed objects in the SAR and HAZMAT community (Breivik and Allen, 2008). The immersion ratio can still be used as an uncertainty parameter to make the projected search area expand at a realistic rate in cases where the object is known to exhibit little or no crosswind motion (Daniel *et al*, 2002), but in many cases the immersion ratio of distressed SAR or HAZMAT object falls within a typical range (for example a life raft with ballast and typical loading) and can for simplicity be assumed to be included in the error estimates stemming from the field experiment where the object was studied.

Conducting field experiments for establishing the drift properties of objects represents a significant cost when developing a data base of search objects. A review of which SAR and HAZMAT objects had been studied by field experiment before 1999 is presented by Allen and Plourde (1999). They made recommendations for 63 categories of objects to be included in SAR planning tools. The previously studied leeway objects were organized into a hierarchical taxonomy based upon their leeway characteristics. This allows the operator to rapidly select a class of objects that best matches the search object for that particular case. If uncertainty exists about the exact nature of the search object, generic classes may be chosen where the drift properties from different studies of similar objects have been lumped together. One example is the generic class for person-in-water based on the drift properties of deceased together with people in survival suits and life jackets. The database is far from complete and should be diversified into a set of regional databases covering the typical drifting objects endemic to the various water bodies around the world oceans.

A large number of SAR and HAZMAT objects were studied a long time ago with methods now considered obsolete. The quality of the field work affects the rate of expansion of search areas when the data are used for predicting the location of a missing object. Hence, tighter confidence limits mean less spread and consequently smaller search areas. Reducing the uncertainty of older object categories and establishing new object categories may thus significantly cut the scope and magnitude of search operations. Field work is expensive and time consuming but crucial to expanding the data base of search objects. It is therefore important to agree on a common procedure for conducting field experiments that allows the results to be easily exchanged between organizations and nations and thus to be utilized by existing and future trajectory forecast models.

This paper is organized as follows. In Section 2 we put forward an operational definition of the *leeway* of an object drifting under the influence of waves and wind. This is essential both to ensure that field data on drifting objects are collected in a coherent fashion and also for guiding modellers in their subsequent effort to choose the most pertinent current vectors when making trajectory forecasts based on the field data. We then assess the uncertainties associated with the main object categories studied so far and categorize the field trials according to which field method was used. We describe our recommended practice for condensing leeway data down to a minimum set of parameters suitable for ingestion in operational search and rescue trajectory models before we propose a method for conducting leeway field measurements in a way consistent with previous field work (and with common practice in SAR trajectory modelling) and define a set of categories of object classes determined by their sizes and their capacity for carrying measuring equipment.

Section 3 illustrates our recommended procedure for collecting and condensing leeway data in a field experiment where three objects of general concern to maritime safety were studied, namely shipping containers, oil drums and drifting World War II mines. We also describe in some detail our choice of instrumentation to highlight the advantages and disadvantages of the different measurement configurations.

## 2. Estimating the leeway of a drifting object

We follow the definition put forward by Allen and Plourde (1999) and state that

> *Leeway is the motion of the object induced by wind (10 m reference height) and waves relative to the ambient current (between 0.3 and 1.0 m depth).*

Establishing an operational definition of the leeway is important for two distinct reasons. First, in order to carry out leeway experiments in a consistent fashion it is important to agree on a standard for measuring the wind and wave-induced response of the object and its motion relative to the ambient water. Second, in order to use the measurements in trajectory models it is important to select the most pertinent wind and current vectors available from numerical models.

The current between 0.3 and 1.0 m depth is roughly coincident with the typical measurement depth of high-frequency (HF) radars (depending on the electromagnetic wave length, but typically 0.5 m; see Fernandez *et al*, 1996 and Breivik and Sætra, 2001) and surface layer drifters. Thus HF radar measurements or surface layer drifters can be used as an alternative to *in situ* current measurements where this proves impractical (especially with smaller objects). It also means that where HF radar measurements are available in real time, short-term current forecasts based on HF current fields (Ullman *et al.*, 2006; Ohlmann *et al.*, 2007) can be used to compute the evolution of search areas.

Likewise, a .standard for wind measurements is important. We follow the meteorological standard and measure the 10-minute vector-averaged wind speed and wind direction. An assumption about the logarithmic wind profile above the sea surface must be made to scale the wind up to 10 m height. Here we follow Smith (1988), but Large *et al.* (1995) is also frequently used.

It is assumed in the leeway definition that the motion through the ambient water results from the joint action of wind and waves. The wind works directly on the over-water structure while waves exert a force on the structure in addition to advection with the Stokes drift. It can be shown (Breivik and Allen, 2008) that wave drift forces on small objects (less than 30 m), such as cargo containers or oil drums, decay rapidly as the ratio of the dominant wave length over the object's length increases and can be neglected compared to wind forces as soon as the wave length is more than about six times the object's length (see also Hogdins and Hodgins, 1998 and Mei, 1989). Hence it can be assumed that for objects even as large as a cargo container or a small boat, leeway can be expressed in most sea states as a function of the wind only. Furthermore, the Stokes drift can under normal circumstances be assumed to be directed along the general wind direction and is confined to a narrow layer near the surface. It is thus practical to have an operational definition of leeway which does not distinguish between the wind and the wave influence.

*2.1. Leeway speed and divergence*

The leeway of small drifting objects tends to increase linearly over typical wind speed ranges. Fitzgerald *et al* (1993) investigated the assumption that the object will rapidly reach its terminal velocity, hence acceleration can be ignored and a simple balance of forces remains (Hodgins and Hodgins, 1998; Breivik and Allen, 2008). Fitzgerald *et al* (1993) also found that the maximum correlation of the leeway occurred at zero lag with 10-minute samples.

Allen and Plourde (1999) compiled the *leeway speed* as linear functions of the wind speed of all 63 object classes studied to date and *divergence angle,* i.e, the angle between the direction of drift of an object and the *downwind* direction during a single sampling period. The use of leeway rate (a percentage of wind speed) and divergence angle was the preferred method of implementing the leeway component of drift in the manual and analytical SAR planning tools.

*2.2. Downwind and crosswind leeway components*

Allen (2005) found that it is better to decompose the leeway into *downwind* (DWL) and *crosswind components of leeway* (CWL) because the downwind component tends to follow an almost linear relationship with the wind speed and this allows an analysis of the crosswind component relationship with the wind separately from the downwind component. The use of downwind and crosswind components of leeway is also better suited for implementation into Monte Carlo SAR planning tools. Some drift objects (usually nearly radial symmetrical) have very little crosswind drift and it may not be possible to discern a clear relationship between the wind speed and the crosswind drift while other objects (e.g. sailboats) have significant right (positive) and left (negative) crosswind components of leeway leading to rapid expansion of two separate search areas (one for right-drifting and another for the left-drifting scenario). Thus, at low wind speeds where the wind direction starts to fluctuate, the variance of the leeway angle increases with decreasing wind speed, making it a non-stationary statistic which is difficult to work with.

An almost linear relationship between the 10-m wind speed $W_{10}$ [m s$^{-1}$] and the leeway of the object is invariably found in field studies (Allen, 2005), albeit with non-uniform (heteroscedastic) spread increasing with wind speed. This allows us to perform a linear regression between the wind speed and downwind and crosswind leeway components. Using this approach, leeway measurements can be condensed down to nine coefficients (Breivik and Allen, 2008),

$$\begin{aligned} L_{\mathrm{d}} &= a_{\mathrm{d}} W_{10} + b_{\mathrm{d}} + \varepsilon_{\mathrm{d}}, \\ L_{\mathrm{c}+} &= a_{\mathrm{c}+} W_{10} + b_{\mathrm{c}+} + \varepsilon_{\mathrm{c}+}, \quad (1) \\ L_{\mathrm{c}-} &= a_{\mathrm{c}-} W_{10} + b_{\mathrm{c}-} + \varepsilon_{\mathrm{c}-}. \end{aligned}$$

Here, the observed downwind leeway (DWL), $L_{\mathrm{d}}$ [cm s$^{-1}$], is related to the wind speed through $a_{\mathrm{d}}$ [%] and the offset, $b_{\mathrm{d}}$ [cm/s] plus an error term, $\varepsilon_{\mathrm{d}}$ [cm s$^{-1}$]. Similar linear regressions can be performed for both the right (+) and left (-) crosswind leeway (CWL) directions, $L_{\mathrm{c}}$ [cm s$^{-1}$], individually, thus allowing left and right-drifting objects to move differently. Assuming a Gaussian error about the linear regression, the three parameters ($\varepsilon_{\mathrm{d}}$, $\varepsilon_{\mathrm{c}+}$ and $\varepsilon_{\mathrm{c}-}$) suffice to account for the error in downwind as well as left and right crosswind components. Finally, the linear regression coefficients can be constrained through the origin (no low-wind offset, $b$=0). Both sets of coefficients should be computed with their associated standard error if the amount of data allows it (see discussion below).

The tendency of objects to tack or *jibe*, that is, to change from one persistent direction of drift relative to the wind direction (e.g. left of downwind) to the opposite is commonly observed. It is necessary to identify these sign changes to split the left and right-drifting events and then compute the left and right crosswind leeway coefficients. As sign changes are generally rare, the simplest method is to visually inspect a progressive vector diagram (PVD) of the leeway relative to the downwind direction (en example of this can be found in Figure 10). The frequency of jibing can only be determined from repeated longer drift studies of a particular drift object configuration. Our present understanding of the influences on jibing is limited, since only a few field experiments have yielded jibing events. Without sufficient data it is difficult to establish a frequency of jibing for that object configuration. Further state changes to the object's condition include capsizing, swamping and tumbling. These generally require long datasets to establish a frequency of occurrence.

*2.3. The indirect method of estimating leeway coefficients*

Objects have been studied using essentially two approaches. The first is the *indirect method*. This method measures the leeway or slip of the object indirectly by tracking the object's drift and subtracting the current based on measurements made by a nearby current meter, drifters or even by visual estimates of the drift of dye patches, drift nets or fields of debris. The effect of wind on drifters was also studied for oceanographic purposes by Kirwan *et al* (1975) and Richardson (1984) while Richardson (1989; 1997) studied the leeway on ships from data on voluntary observing ships (VOS). The indirect method was the sole technique used to infer the leeway of drifting objects until the early 1990s due to its relative simplicity, with the notable exception of Suzuki and Sato (1977). Early work started with Chapline (1960) and references to later field work can be found in Allen and Plourde (1999). The indirect method was still in use up until the late 1990s (see Fitzgerald *et al,* 1990 and further references found in Allen and Plourde, 1999). Although straight forward in its application, the method is prone to errors as the object must be kept close to the current meter, a difficult task especially in heavy weather.

*2.4. The direct method of estimating leeway coefficients*

Later, as lighter current measuring devices with internal recording became available, the *direct method* came into use. Here, a current meter is towed behind or attached directly to the object of interest. This method is of course well known and extensively used in physical oceanography, with early work by Niiler *et al.* (1987) and Geyer (1989). O'Donnell *et al.* (1997) also followed this approach to studying the slip (leeway) of surface drifters. Larger objects are also capable of carrying a light wind anemometer without seriously changing its over-water structure. This led to huge improvements in the accuracy of the leeway estimates, as is evident from the error estimates listed by Allen (2005). The method came into regular use in the 1990s (see Allen, 1996; Allen and Plourde, 1999 and Turner *et al*, 2006).

*2.5. Limitations to the estimation of leeway drift properties*

Figure 1 shows the leeway divergence (deviation from the vertical axis) and leeway speed (distance from the origin) of a selection of objects studied to date. The non-ballasted life rafts and the fishing vessels represent older object categories studied using the indirect method while the more modern deep ballast life rafts were studied with direct measurements of the leeway using attached current meters. It is evident that the experimental spread (an estimate of the spread about the regression between wind speed and leeway estimate) goes down dramatically when direct leeway measurements are made. It is important to note, however, that this is a very limited assessment of the measurement error as it only measures deviations from the straight line regression. But, other things being equal, this gives an estimate of the potential for reducing the measurement error by employing direct measurements.

The method outlined above for estimating leeway coefficients depends on a certain amount of leeway measurements (typically 10-minute vector averages of wind and leeway vectors) having been successfully collected. It may not be possible to do a full analysis with unconstrained and unsymmetrical leeway coefficients if the field data are limited in time and/or the wind conditions show little variation. Analysis methods for leeway data will then depend upon the quantity and quality of the leeway and wind data as follows.

1. **Constrained, symmetrical regression ($b$=0, $a_{c+}=a_{c-}$):** When the data set includes accurate measurements of leeway and wind direction, but the range of wind speed is limited, only constrained linear regression may be calculated for the downwind (DWL) and crosswind components of leeway (CWL) versus wind speed. Since the object can drift to the left or to the right of the downwind direction, we assume that CWL is symmetrical about the downwind direction when fitting regression of CWL versus wind speed.
2. **Unconstrained, symmetrical regression ($a_{c+}=a_{c-}$):** When data are collected over a range of wind speeds of typically 1-10 m $s^{-1}$ and the number of data samples $N$ exceeds approximately 50 both constrained and unconstrained linear regressions of DWL and CWL can be calculated.
3. **Unconstrained, unsymmetrical regression:** When the data set includes multiple drift runs or jibing events in a single, long drift run (e.g., $N$>100), the CWL data can be split into left and right-drifting segments by visual inspection of the PVDs. Now separate CWL coefficients for left (negative) and right (positive) of downwind can be estimated and it is no longer necessary to assume symmetry.
4. **State changes:** With longer datasets ($N$>500) and multiple runs with several jibing events covering wide ranges of wind conditions it may become possible to fully characterize the behavior of the leeway object by estimating the frequency of events of jibing, capsizing and swamping. An average probability of jibing of 4% per hour has been estimated by Allen (2005) based on a compilation of PVDs from a range of small objects. We have no estimates of the frequencies of state changes such as capsizing and swamping, but observations of a cap-sized six-person liferaft have been reported (Allen and Fitzgerald, 1997).

*2.6. Experimental drift objects classified by size*

As ship time and equipment are limited commodities it is tempting to pool together different objects for the same field experiment. But there are also practical reasons for doing so; certain objects may be too small to carry the instrumentation required to measure the local wind. Under certain conditions the simultaneous release of small objects and larger objects carrying a wind anemometer can allow dual use of the wind measurements as the wind varies quite gradually in offshore locations. If the object is also too small to carry or tow a current meter, the indirect method must be used and the current must be derived by some other means, for example from land-based HF radars, Lagrangian surface drifters or an Eulerian current meter in a fixed position nearby.

We define four sizes of drift objects, depending upon the kind of instrumentation the object can accommodate.

1. The smallest drift objects may only be able to carry a tracking device. These objects can only be studied using the indirect method where their leeway components are determined by subtracting the surface current portion of drift from the total displacement over the sampling periods. Typical objects of this size include Emergency Position-Indicating Radio Beacons (EPIRB), see Turner et al (2006) and medical syringes and vials (Valle-Levinson and Swanson, 1991).
2. Objects large enough to accommodate a tracking device and a current meter but too small to be equipped with a wind anemometer constitute the second category. These allow for determination of leeway using the direct method. Simulated persons-in-water, sea kayakers (see Allen *et al.*, 1999), and in this study, a 220 l (55-gallon) oil drum, are examples of this category.
3. Mid-sized drift objects can carry tracking devices, a current meter and a meteorological package for measuring the local surface winds. Life rafts, small boats, and in this study a down-scaled 1:3.3-sized model of a 40-ft shipping container are examples of mid-sized drift objects.
4. Large objects can also house the meteorological instrumentation package but special consideration must be taken since they are either vessels with persons on board or are too large to be easily deployed and recovered over the side of the working vessel. Conducting leeway drift studies on a vessel with people onboard is limited by the stamina of the crew on board. Large objects without persons on board, such as 20 or 40-ft shipping containers, are limited by the ability of the research vessel to deploy and recover or tow the objects to and from the test area. These objects may also have draft considerably deeper than 1 m and therefore be affected by deeper currents in the presence of vertical shear; larger momentum (non-zero lag with the winds) and direct wave forcing. To date none of these effects have been adequately studied.

# 3. A field study of the drift properties of three objects

Four separate drift runs were conducted on three drift objects from 31 March to 4 April 2008 with the support of the Norwegian Coast Guard (Kystvakten) vessel K/V Ålesund off Fedje, Norway (located at 60.5ºN, 004ºE, see Figure 2). The location was close to the home port, ice free, and experienced a wide range of wind conditions over the four-day trial period. The Norwegian Coastal Current flows northward through the experiment area at an average speed of approximately 50 cm $s^{-1}$ (Breivik and Sætra, 2001, Essen *et al*, 2003). A high-frequency (HF) radar network covers the region and provided valuable independent current estimates throughout the field campaign.

Wave measurements were obtained from an Anderaa ADCP moored on site for the duration of the field tests. Over the whole campaign, the sea-state was moderate, with a significant wave height of about 2.3 m and peak periods in the range 6.5 to 9.5 s. We thus consider the wave drift forces negligible for our test objects.

Each object was equipped for deployment, drift and recovery and data collection. The objects were ballasted and augmented with extra flotation to mimic their distress configuration as closely as possible.

Measurement records were matched in time for analysis on 10-minute vector averages of leeway and wind. The leeway was decomposed into downwind and crosswind components for every 10-minute sample. The downwind leeway component was linearly regressed against the wind speed adjusted to 10 m. The linear regression was done both unconstrained and constrained through the origin. The slope, the offset ($y$-intercept; thus only for the unconstrained regression), the root-mean-square of the regression residuals, $S_{yx}$, and $r^2$ were computed following Neter *et al* (1996). For the crosswind components of leeway, the values were separated along runs or portions of runs indicated by the progressive vector diagrams (PVD) to be consistently left (negative) or right (positive) of the downwind direction (see Figure 10). Then the crosswind components, negative or positive were linearly regressed against the 10-m wind speed, again both unconstrained and constrained through the origin.

### *3.1. Instrumentation*

#### 3.1.1 Current Measurements

The size and configuration of the underside of the drift object will influence the choice of the current meter. For the 1:3.3-size model container and the WWII mine two InterOcean S4EMCM current meters were used. The model container belongs to our Class 3 of Section 2.6; objects large enough to hold their own equipment. The current meter was placed in an aluminum frame suspended just below a surface float attached by a 30-m line to the pull point on the drift object (see Figure 3). This allows the current meter to be well away from flow distortion under and around the drift objects. For the oil drum an RD Instruments high-frequency acoustic Doppler current meter (ADCP) was used. The oil drum belongs to our Class 2; objects that are too small to carry their own wind anemometer mast. The oil drum has significantly lower flow distortion than the model container but also lower weight. We considered the object too light-weight to be towing a current meter while at the same time also distorting the flow less and decided to build the ADCP into the oil drum flush with its underside rather than towing the current meter behind (see Figure 4).

Additional current measurements were provided by the high-frequency radar network operated by the Norwegian Coastal Administration for the Fedje area (Figure 2).

#### 3.1.2 Measuring the wind field

Mid and large-sized leeway objects (classes 3 and 4 of Section 2.6) can carry meteorological measuring equipment. The primary purpose is to measure the local surface winds, although other useful parameters may include air temperature, water temperature and pressure. The model container was equipped with an R.M. Young Weatherpak anemometer mast. The oil drum and the mine were not large enough to hold a wind anemometer and consequently the wind measurements collected on the model container had to be used. Wind vectors were adjusted to 10 m height following Smith (1988) after being corrected from relative to absolute winds.

#### 3.1.3 Tracking and recovering objects

During this experiment, Class-B Automatic Identification System (AIS) transponders were used to track and recover the drift objects. The AIS consists of a very high frequency (VHF) transponder and receiver attached to a GPS, broadcasting a signal every 2 to 10 seconds depending on vessel speed. For the experiment, three Class-B AIS transponders were built as self-contained units consisting of an AIS unit, a 24 Ah battery, a GPS antenna, a VHF antenna and a data logger. The AIS transponder not only assisted in tracking and recovering the objects, but also helped make the objects more visible to traffic in the vicinity of the experiment.

Argos, VHF, and strobe-flasher beacons were all used to aid in the tracking and recovery of the drift objects. GPS data loggers provide speed and course over ground. We have used pairs of mercury-switch Argos beacons, one oriented upright and one downward, to provide tracking while the object was upright and also in case it would capsize. Small VHF transmitter and flashers (Novatech beacons) were attached using tag lines, these provide tracking for both upright and capsized drift objects.

### *3.2. Scaled-down model of 40-ft shipping container*

A scaled-down model of a full 40-ft container was designed for this experiment as the coast guard vessel did not have the equipment required to handle a full-size shipping container. Studying the drift of a medium-sized container is valuable in itself as its dimensions and loading make it quite representative of smaller storage containers. It also provides interesting data for studies of scaling effects on drift damping. Daniel *et al* (2002) studied a 20-ft shipping container under various loading conditions. However, their study failed to assess the crosswind leeway component and to establish confidence limits on the leeway coefficients. Hence, the objective of this study was both to revisit the leeway estimates of Daniel *et al* (2002) and to assess the cross wind component of shipping containers.

The model container is depicted in Figure 3 with the wind anemometer mounted and the current meter towed behind. The object was immersed to approximately 70% to mimic the typical loading of real shipping containers. The linear regression slope, offset, correlation, $r^2$ and standard error ($S_{yx}$) for the leeway speed, downwind component of leeway and crosswind components of leeway are summarized in Tables 1 and 2. The downwind and crosswind leeway components are shown in Figure 7 and Figure 8 along with the unconstrained and constrained linear regressions and their respective 95% confidence limits. It is clear from Figure 8 that the crosswind component is negligible for the container.

Daniel *et al*. (2002) followed the same analytical approach as Geyer (1989) to estimate the leeway ratio (leeway speed / wind speed) of a shipping container as a function of immersion rate. Following Geyer (1989) this can be formulated as

$$\frac{u_{\text{leeway}}}{u_{\text{wind}}} \sim \sqrt{\frac{\rho_{\text{air}} C_{\text{D}} A_{\text{air}}}{\rho_{\text{water}} C_{\text{W}} A_{\text{water}}}} \qquad (1).$$

Here, the density of air and water are assumed to be $\rho_{\text{air}} = 1.29$ kg m$^{-3}$ and $\rho_{\text{water}} = 1025$ kg m$^{-3}$ and the area of the over-water structure and the submerged parts of the object are given by $A_{\text{air}}$ and $A_{\text{water}}$. Furthermore, $C_{\text{D}}$ and $C_{\text{W}}$ are the drag coefficients on the over-water and submerged parts of the object, respectively. This can be rewritten in terms of an immersion ratio

$$I = \frac{A_{\text{water}}}{A_{\text{air}} + A_{\text{water}}}$$

to read

$$\frac{u_{\text{leeway}}}{u_{\text{wind}}} \sim \sqrt{\frac{\rho_{\text{air}} C_{\text{D}}}{\rho_{\text{water}} C_{\text{W}}} \frac{1-I}{I}} \qquad (2).$$

We found the leeway-to-wind ratio to be approximately 1.4% for the model container immersed to around 70% ($I$ = 0.7), using unconstrained regression and a ratio of 2% when using the constrained regression (see Tables 1 and 2). This agrees reasonably well with Eq (16) of Daniel *et al* (2002) where a leeway-to-wind ratio of about 2.2% is found. The main reason for the higher theoretical leeway ratio found by Daniel *et al* (2002) is probably their choice of drag coefficients ($C_{\text{D}} = C_{\text{W}} = 1.0$) which was obtained from tank tests on models that do not take into account the heave, pitch and roll of open ocean conditions which induce additional viscous damping. Furthermore, a $C_{\text{D}}$ of 1.0 corresponds to a box shape with one face perpendicular to the air flow. Considering the box with its edge in the wind (a feature which was observed during the experiment), $C_{\text{D}}$ can be reduced to 0.8. By also increasing $C_{\text{W}}$ to 1.2, Eq (2) yields a leeway-to-wind ratio of 1.9. It is also clear that the smaller vertical extent of the scaled-down container made it more sheltered by the waves, but without more detailed measurements of the wind profile at several levels it is difficult to establish to what extent this effect made a significant difference when comparing with the full 20-ft container. The observed discrepancies between the drag coefficients applied by Daniel *et al* (2002) and those inferred from the leeway to wind ratio of our scaled-down container are at any rate within the range of experimental error in our field experiments.

We also see that the container exhibits very little crosswind drift, in line with the assumptions of Daniel *et al.* (2002). It is obvious that a scaled-down model container will not behave identically to a full-scale 40-ft container, but the match with the results found by Daniel *et al* (2002) is promising, indicating that their results can be used for a range of container-like objects. Further work is required to collect leeway measurements on full-scale 20 and 40-ft shipping containers. Several immersion levels, especially 10-20% (empty containers) and 80-95% (fully loaded containers) should be studied to further assess the curves of Daniel *et al.* (2002).

*3.3. 220 l (55-gallon) Oil drum*

A 220 l (55-gallon) oil drum is shown in Figure 4. An ADCP was fitted inside the drum looking down. This configuration was chosen instead of towing because the drum is so light that a towed current meter might seriously affect its motion. The disadvantage is that flow distortion around the underside of the drum may affect the current measurements. The drum was also too small to host its own wind anemometer mast. The downwind component of leeway as a function of 10-m wind speed is shown in Figure 9 along with the unconstrained and constrained linear regressions and their respective 95% prediction limits. Tables 1-3 summarize the linear regression slope, offset, $r^2$ and standard error for the leeway speed, downwind component of leeway and crosswind components of leeway. The crosswind components of leeway for the drum were split into positive and negative values after inspecting the progressive vector diagram of the downwind and crosswind components of leeway displacement (Figure 10). The two major sign changes (jibes) in crosswind component from negative (drift left of downwind direction) to positive (right of downwind) are indicated by arrows in Figure 10. The left and right-drifting leeway coefficients are shown in Table 3.

Most of the total displacement for the four drift runs with the oil drum was due to the strong northward-flowing coastal current. HF radar measurements (Figure 2) indicated that the inshore portion of Norwegian Coastal Current was between 0.25 and 1.0 m $s^{-1}$. Although it is important to keep track of the total current to correctly estimate the relative wind speed, it will not affect the measurements significantly as only the relative motion of the object through the water is of relevance.

*3.4. World War II L-MK2 Mine*

A World War II L-MK2 mine was used in this experiment (see Figure 5 and Figure 6). The mine had an oval shape and measured 125 cm in height and had a diameter of 105 cm. The total weight was around 300 kg including 100 kg of sand in replacement for explosives. Roughly 75% of the mine was submerged when it floated freely in the water. An S4 current meter was towed behind the mine with about 30 m of line using the same arrangement as for the container. No wind anemometer was attached to the mine.

The downwind component of leeway relative to the 10-m wind speed is shown in Figure 11, along with the unconstrained and constrained linear regressions and their respective 95% prediction limits. The mine was found to have very little crosswind leeway and a downwind leeway component of approximately 2% of the wind speed. Again Tables 1 and 2 summarize the linear regression slope, offset, $r^2$ and standard error for the leeway speed, downwind component of leeway and crosswind components of leeway.

*3.5. Uncertainty of the leeway estimates*

As seen in Figure 12 the objects cluster quite well in the two-dimensional leeway-space introduced in Figure 1. This is not surprising as all three objects are heavy and well immersed in their distress configuration. The objects exhibit comparatively low error ellipses which is unsurprising as their leeway was measured directly. It is however illustrative of the amount of uncertainty that enters a search at the moment of the field method chosen.

# 4. Conclusion

We have described a standardized method for performing leeway field experiments on SAR and HAZMAT objects. As the objects vary widely in size and shape, it is necessary to allow for some flexibility in the experimental setup. The methodology lays down a rigorous definition of the term *leeway*, which, taken together with standardized wind measurements (10-minute vector averages adjusted to 10 m above the sea surface) allow exchange of field data between different trajectory models. Furthermore, employing a linear relation between the downwind and crosswind components of the leeway along with estimates of the experimental measurement error allows the field results to be condensed down to a set of nine coefficients. Recommendations are made as to the amount of field data required to make reliable estimates of these leeway coefficients. This does not mean that a linear relation must be obeyed, and indeed we recommend that the raw experimental data also be made available, especially as future trajectory models may come to include wave effects.

Leeway data was collected for three drift objects: a 1:3.3 sized model of a 40-ft shipping container model 70% immersed, a 220-l (55-gallon) oil drum, and a floating WWII L-MK2 mine. The shipping container drifted at approximately 1.4% of the 10-m wind speed while the oil drum drifted at 0.8% and the mine at 2%. The mine and the container exhibited negligible crosswind leeway, while the drum had a crosswind leeway of 0.4-0.6% of the wind speed. It is important to note that the constrained regression presented in Table 2 yields somewhat higher leeway-to-wind ratios as there is no offset for zero wind. This makes sense physically as we expect objects to remain at rest relative to the ambient water under calm conditions. However, there are good “operational” reasons for using the unconstrained regression in Table 1 as the error associated with a constrained regression line must be higher and thus the search areas will expand faster using the figures in Table 2 under moderate wind conditions

The method outlined in Section 2 for organizing field experiments for several leeway objects was followed and was found to work well for our three objects. Relying on the wind anemometer on the model container was found to work in our case because the winds were quite uniform throughout the domain and because the objects kept relatively close. Towed current meters seem superior to gimbled ones as they tend to be less influenced by the roll and pitch of the object itself. Also, flow distortion around the underside of the objects may cause deviations from the real motion through the water.

The results presented include objects varying in size and shape from a standard oil drum to a scaled-down 40-ft container. The leeway of the objects show the same linear relationship found also in earlier studies. The experimental spread is however quite extensive, suggesting that there is potential for future experimental setups which relate the residual variance to other geophysical parameters, in particular to wave energy travelling in off-wind directions (swell). It is unclear to what extent waves influence the frequency of jibing of a drifting object. To establish this would require time series of significant wave height and possibly a full two-dimensional spectrum.

The loading of the objects was deliberately chosen to coincide with what we consider the likely distress configuration of such objects when they become a search object. It is however clear that as the loading must vary greatly for such objects, such experiments should be carried out also for different loading conditions. This may not be so for all objects, though. Sail boats, persons in water and large life rafts may reasonably be assumed to have more uniform distress configurations than for example containers and oil drums. There is no final answer to how objects of variable loading should be handled in an operational setting, but the most straight-forward way is probably to include different ranges of loading.

The experiments were carried out in an area where the Norwegian Coastal Current is quite strong and currents exceeding 1 m $s^{-1}$ were observed during the field trial in the HF radar current maps, as is clearly seen in Figure 2. As expected, the strong currents seem not to have affected the measurements significantly as the measurement spread is comparable to what has been found in earlier leeway studies (Allen, 2005). This is to be expected as all measurements of wind and leeway are done relative to the drifting object. It is however recommended to have independent current measurements in the form of HF radar, drifters or Eulerian current meters to monitor the strength of the surface current.

The leeway estimates of the three objects (Figure 12) were found to have low error ellipses compared to older leeway categories studied using the indirect method. This is unsurprising but poignantly reminds us that the success of a search operation relies crucially on the quality of the field work that went before.

*Acknowledgments*
The authors would like to acknowledge the substantial support that this work has received through the ship time allotted by the Norwegian Coast Guard, without which there would be no field work. The work was conducted as part of the SAR-DRIFT project, funded by the French-Norwegian Foundation (Fondation Franco-Norvegienne) and Eureka project E!3652. The Norwegian Joint Rescue Coordination Centres and the Norwegian Navy have contributed to the development of the leeway field techniques through a number of grants over the past 10 years.

# List of Figures

Figure 8. Crosswind component of leeway (cm s-1) versus wind speed adjusted to 10-meter height for the 1:3.3 sized 40-foot Shipping Container. Unconstrained linear regression mean (solid) and 95% confidence levels (dash) are plotted in green. Constrained linear regression is plotted in red. It is evident that cross wind drift is very weak.

Figure 9. Downwind Component of Leeway (cm s-1) v wind speed adjusted to 10-m height for the 220-l (55-gallon) oil drum. Unconstrained linear regression (solid) and 95% confidence levels (dash) are plotted in green. Constrained linear regression is plotted in red. Run 1 data are plotted in blue, Run 2 in red, Run 3 in green and Run 4 in black.

Figure 10. Progressive Vector Diagram (PVD) of the downwind and crosswind components of leeway displacement, for the 220-l (55-gallon) drum. Downwind direction is up, positive crosswind to the right, displacement is in kilometers. Hourly intervals are indicated by a hatched point every 6th point. Horizontal black arrows indicate the two major shifts in the sign (jibes) of the crosswind component of leeway that were found in the four runs.

Figure 11. Downwind component of leeway (cm s-1) v wind speed adjusted to 10 m height for the WWII L-MK2 mine. Unconstrained linear regression mean (solid) and 95% confidence levels (dashed) are plotted in green. Constrained linear regression is plotted in red.

Figure 12. Distribution of the three objects studied in the two-dimensional leeway-space (red) at 10 m s-1 wind speed. The object categories from Figure 1 are reproduced in black for comparison. The distance from the origin indicates leeway speed [cm s-1] while the angle relative to the vertical indicates the object's divergence from the wind direction (wind blowing upwards along second axis). The ellipses show crosswind and downwind error. Categories are placed selectively in the left and right quadrant for readability.

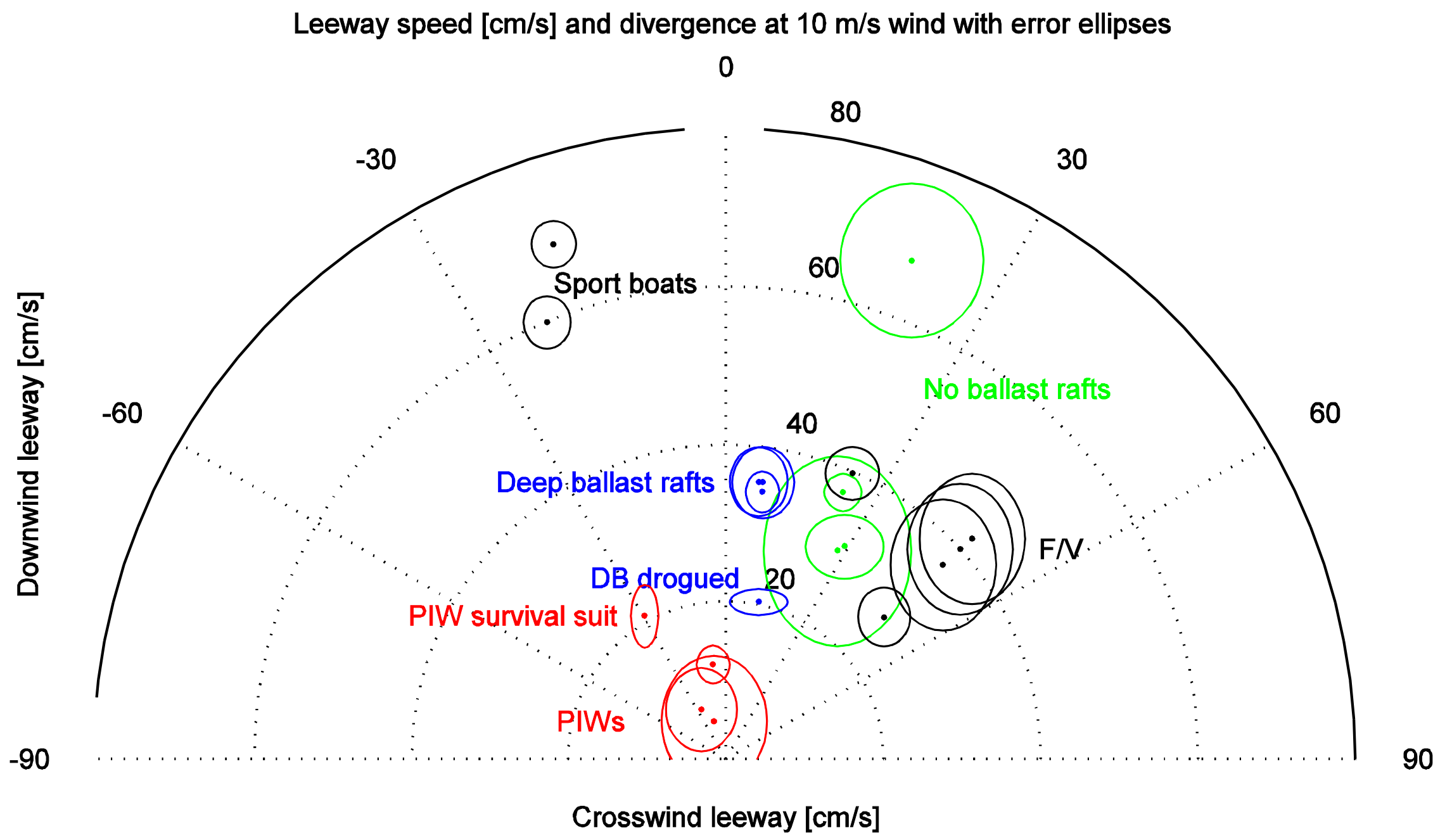

Figure 1. Distribution of selected object categories in the two-dimensional leeway-space. The downwind (vertical axis) and crosswind (horizontal axis) leeway of selected leeway categories at 10 m/s wind speed is shown. The distance from the origin indicates leeway speed [cm/s] while the angle relative to the vertical indicates the object's divergence from the wind direction (wind blowing upwards along second axis). The ellipses show crosswind and downwind error. Categories are placed in the left and right quadrant for readability. Deep ballast rafts (DB, marked blue) and the person in water (PIW) in survival suit (red) are objects studied with the direct method. The older unballasted liferafts (green) and fishing vessels (F/V, black) were studied earlier using the indirect method.

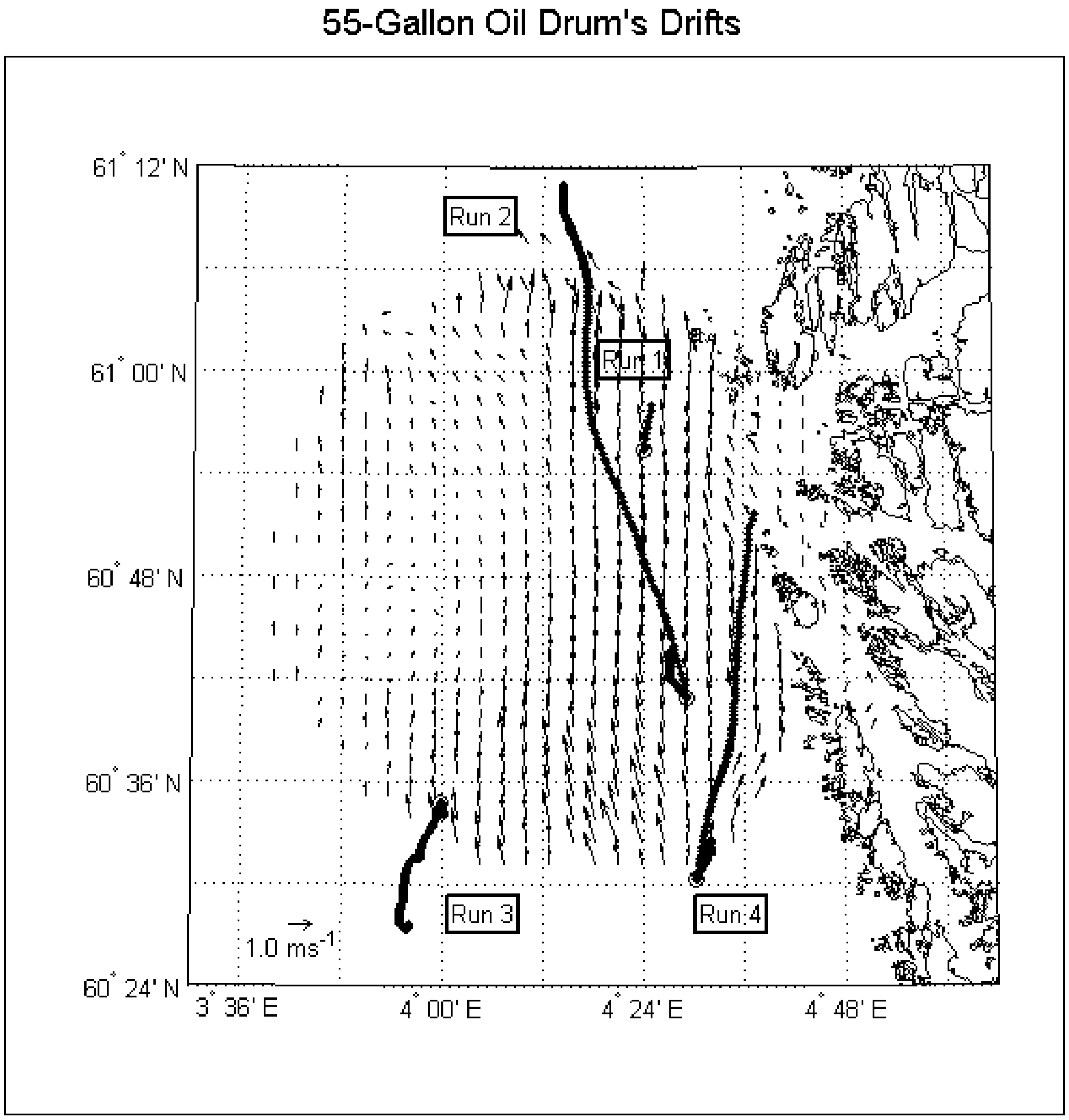

Figure 2. HF radar current map valid for 2008-04-01T03:00 UTC with the trajectories from the four runs with the oil drum overlaid in red. The Norwegian Coastal Current exceeded 1 m $s^{-1}$ during the field experiment. This is made evident by the small wind-driven (leeway) contribution to the total trajectories. The four runs were deployed between 31 March and 3 April with final pickup 4 April.

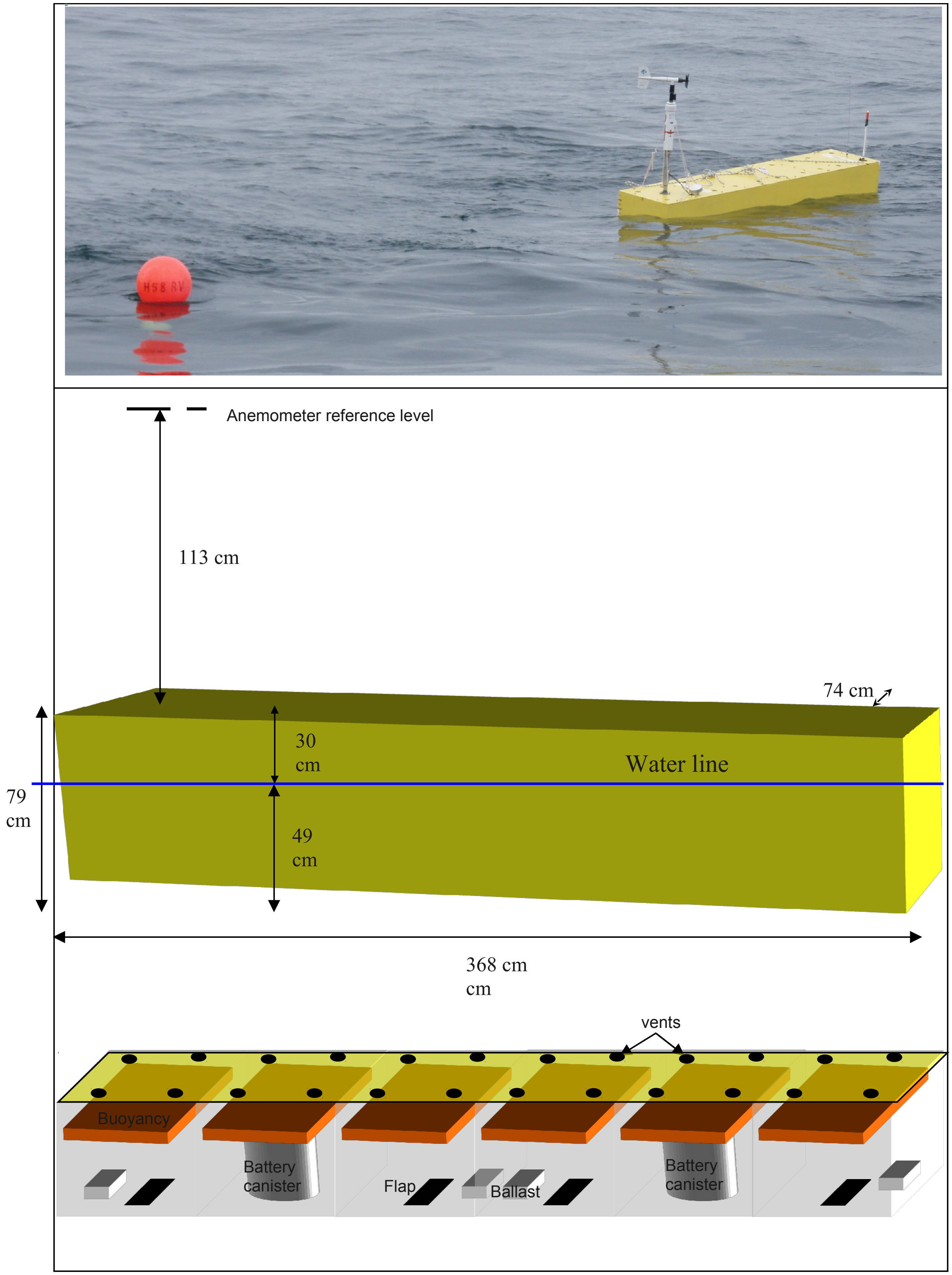

Figure 3. Photograph, three-dimensional line drawing and sketch of internal arrangement of a downscaled (1:3.3) model of a 40-ft shipping container immersed to 68%. The model is equipped with a WeatherPak anemometer system and S4 current meter towed (just visible beneath the orange float in the photograph). The underside of the container has flaps that open to let water in as the object is lowered into the ocean.

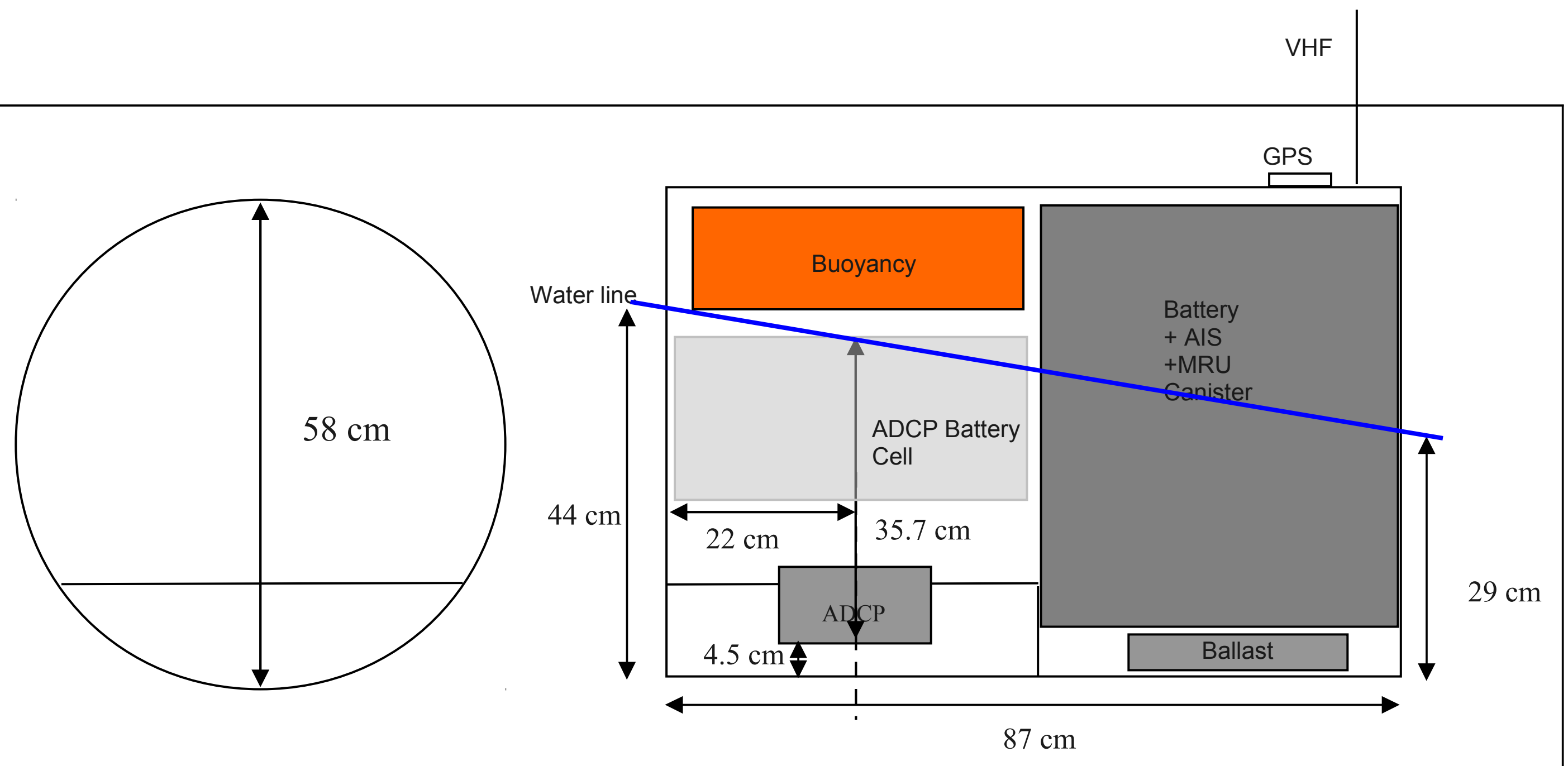

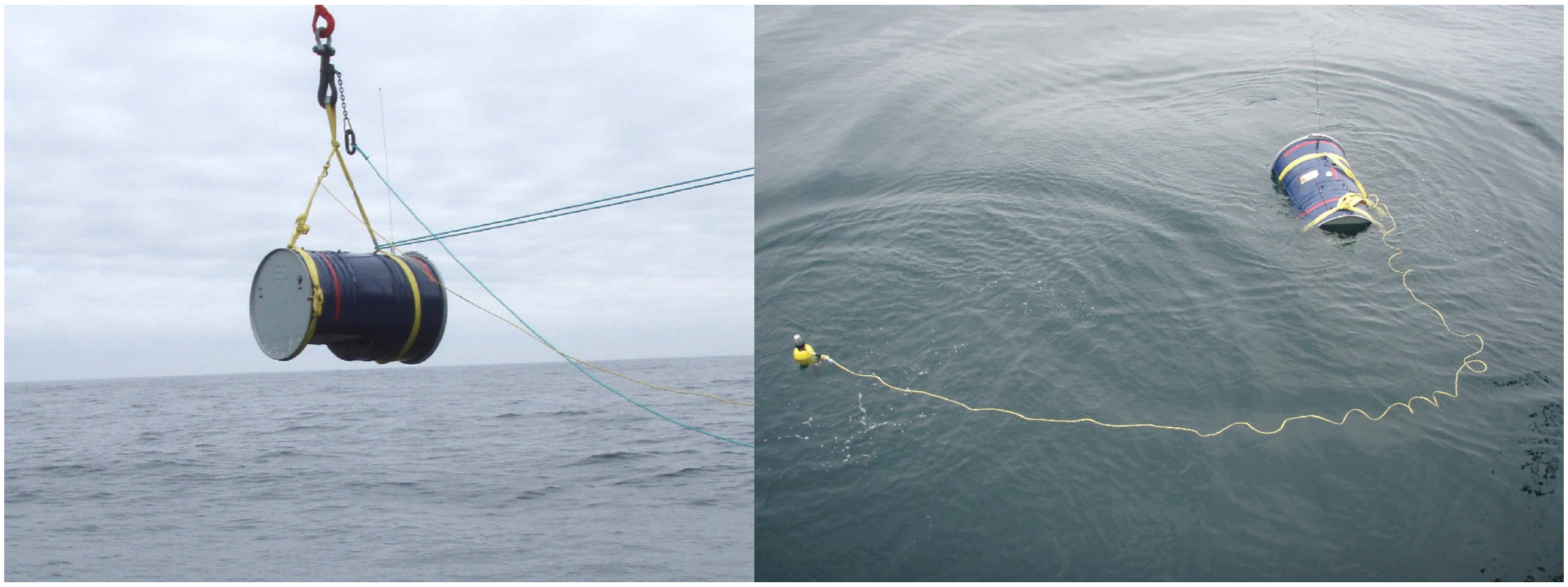

Figure 4. The oil drum used for the trials is a 220 l (55-gallon) steel drum. An aperture in its side allows the gimbled downward-facing ADCP inside to be mounted flush with its surface. The AIS, MRU, data recorder and batteries are placed in a watertight canister which occupies 30% of the drum's volume. Additional buoyancy volumes and 2 kg of ballast are added so that the drum floats horizontally with a slight tilt of about 10° with the horizontal and an average draught of about 63% of its diameter. VHF and GPS antennas for the AIS transponder are flush mounted on the side of the drum above the AIS canister in order to minimize submergence.

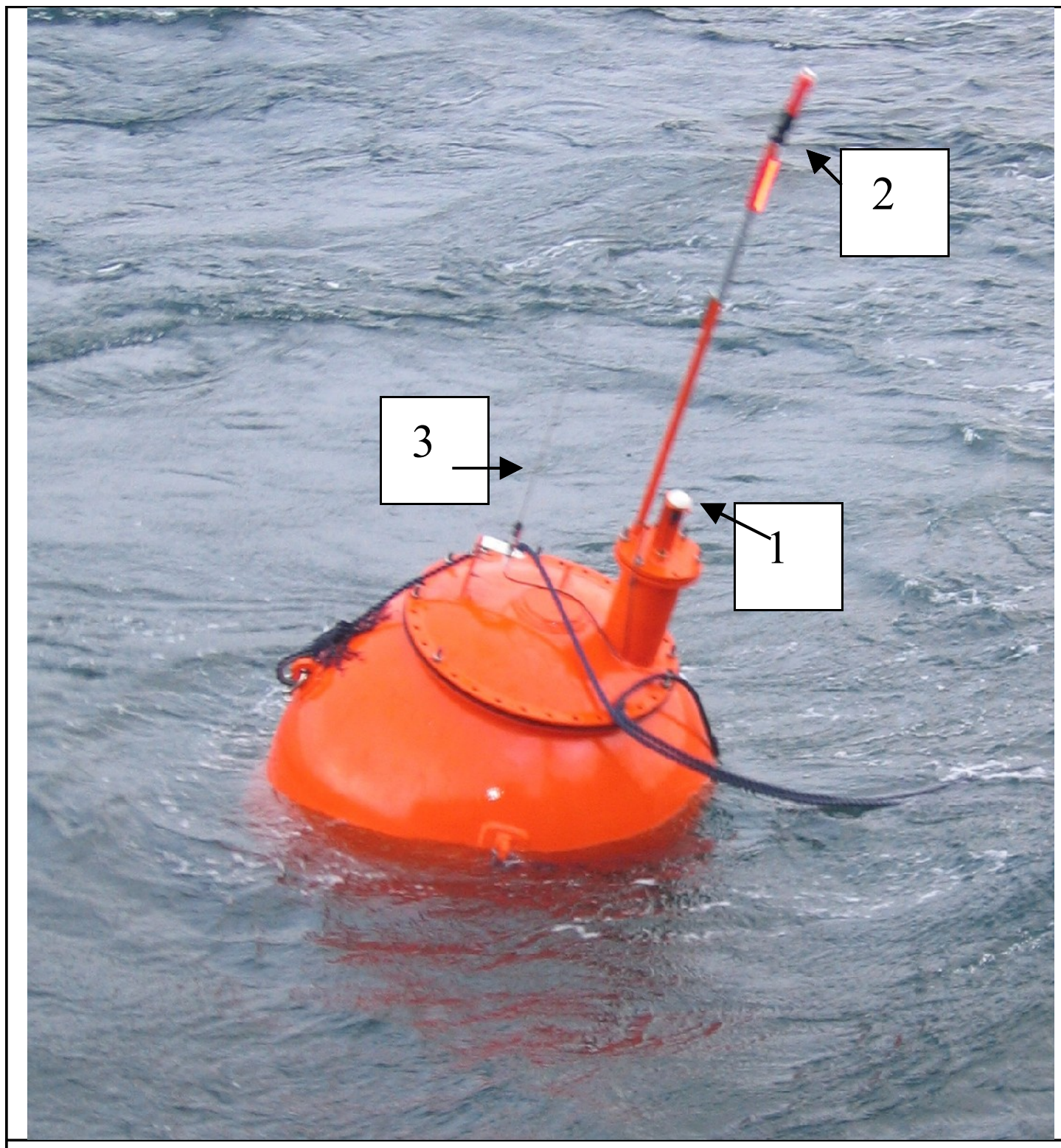

Figure 5. Photograph of World War II L-MK2 mine. Roughly ¾ of the mine was submerged when it floated freely in the water. Attached to the lid is a signal light (1), a GPS antenna (2) and VHF antenna (3). The AIS unit is placed inside the mine.

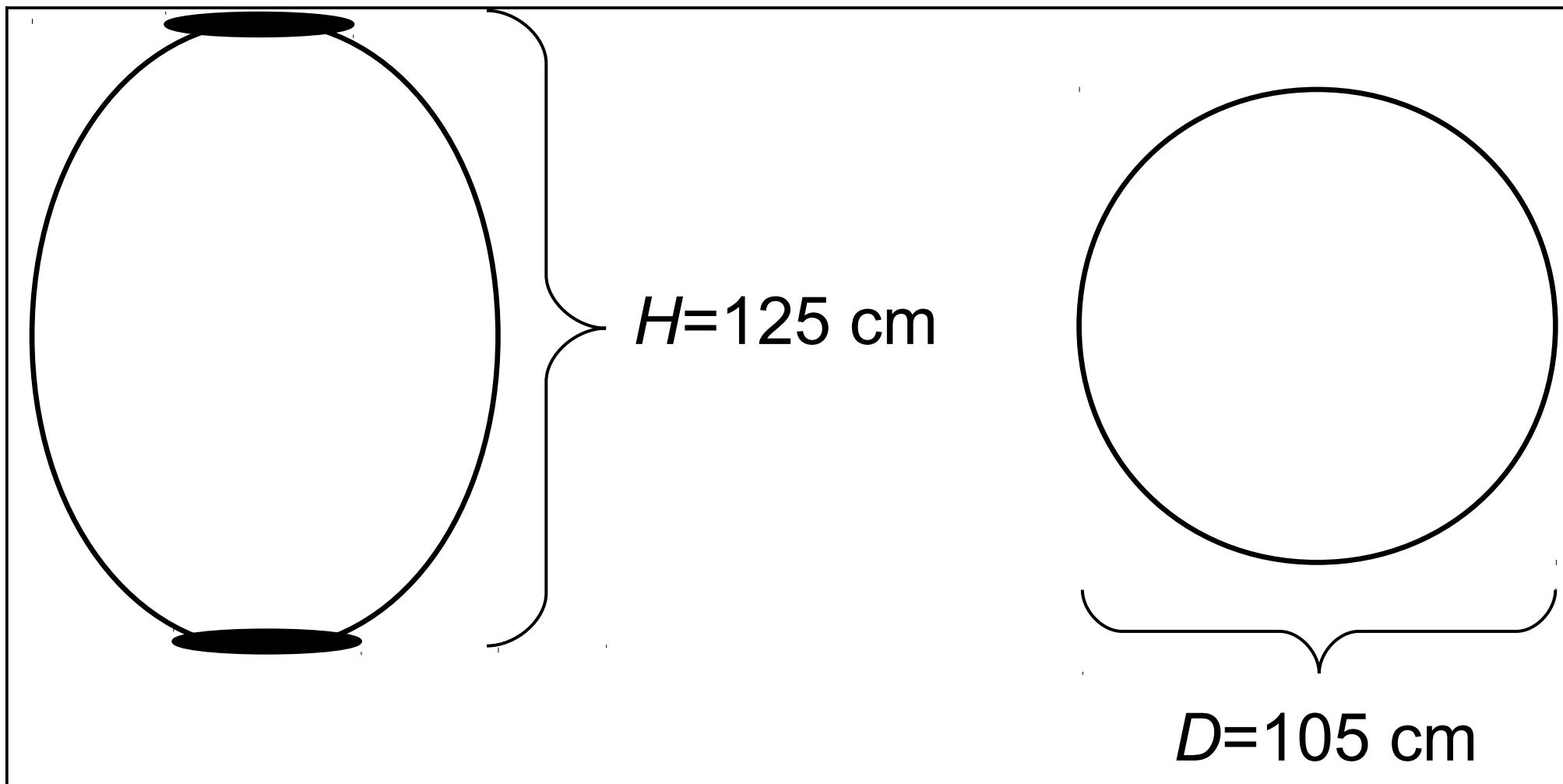

Figure 6. Line drawing of World War II L-MK2 mine. The mine measured 125 cm in height and had a diameter of 105 cm. It contained 100 kg of sand in replacement for explosives and had a total weight of 300 kg.

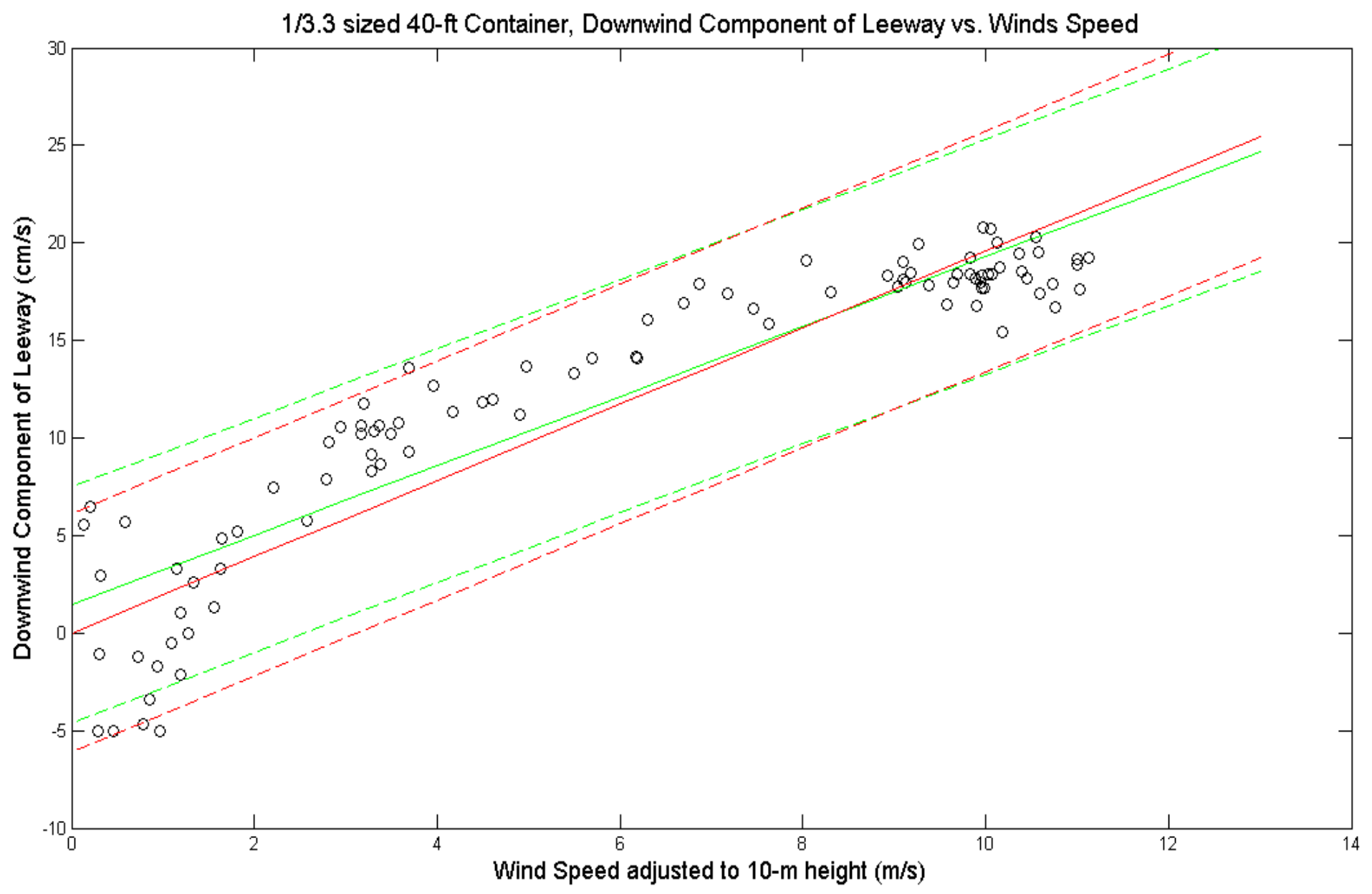

Figure 7. Downwind Component of leeway (cm $s^{-1}$) versus wind speed adjusted to 10-meter height for the 1:3.3 sized 40-foot Shipping Container. Unconstrained linear regression mean (solid) and 95% confidence levels (dash) are plotted in green. Constrained linear regression is plotted in red.

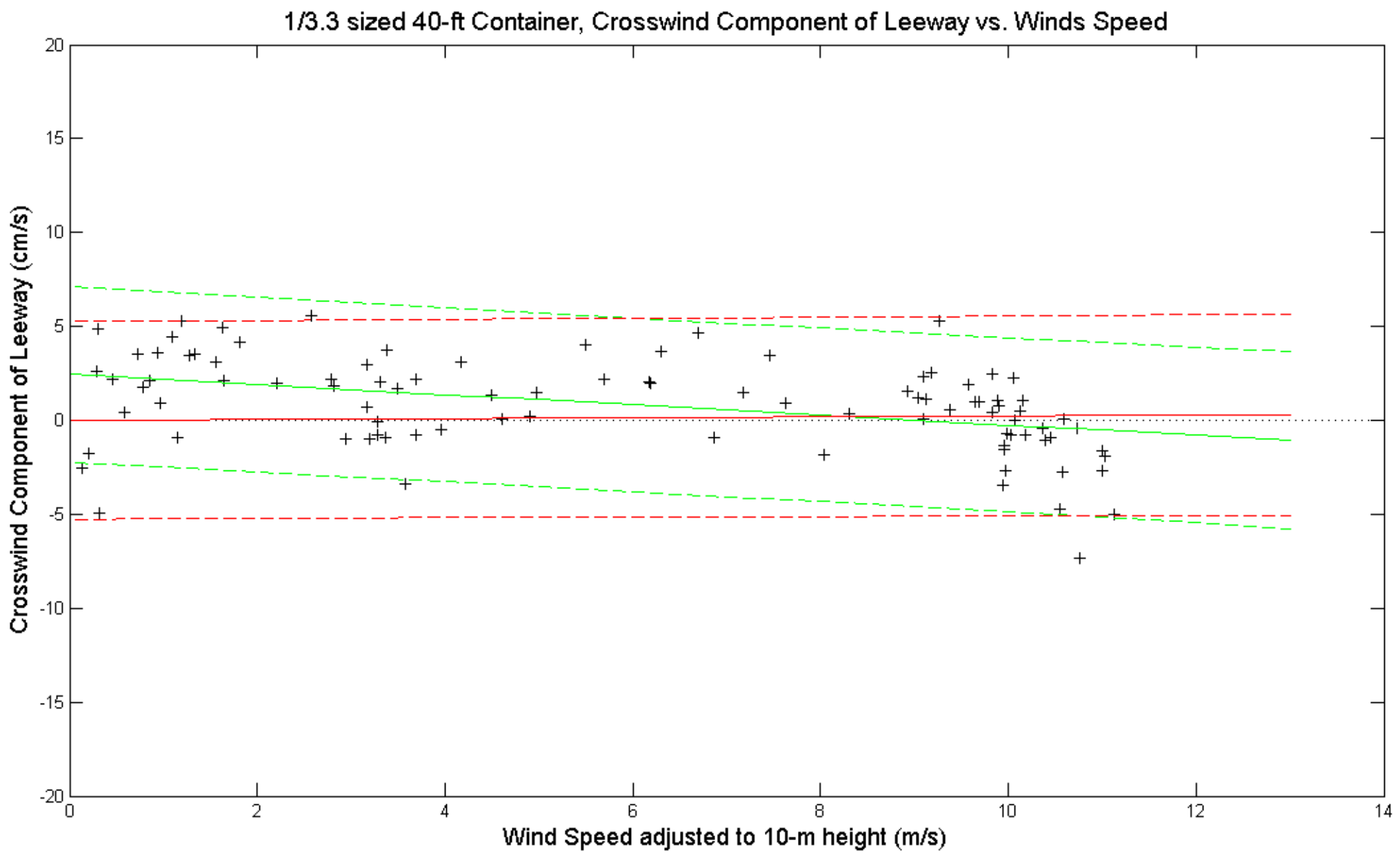

Figure 8. Crosswind component of leeway (cm $s^{-1}$) versus wind speed adjusted to 10-meter height for the 1:3.3 sized 40-foot Shipping Container. Unconstrained linear regression mean (solid) and 95% confidence levels (dash) are plotted in green. Constrained linear regression is plotted in red. It is evident that cross wind drift is very weak.

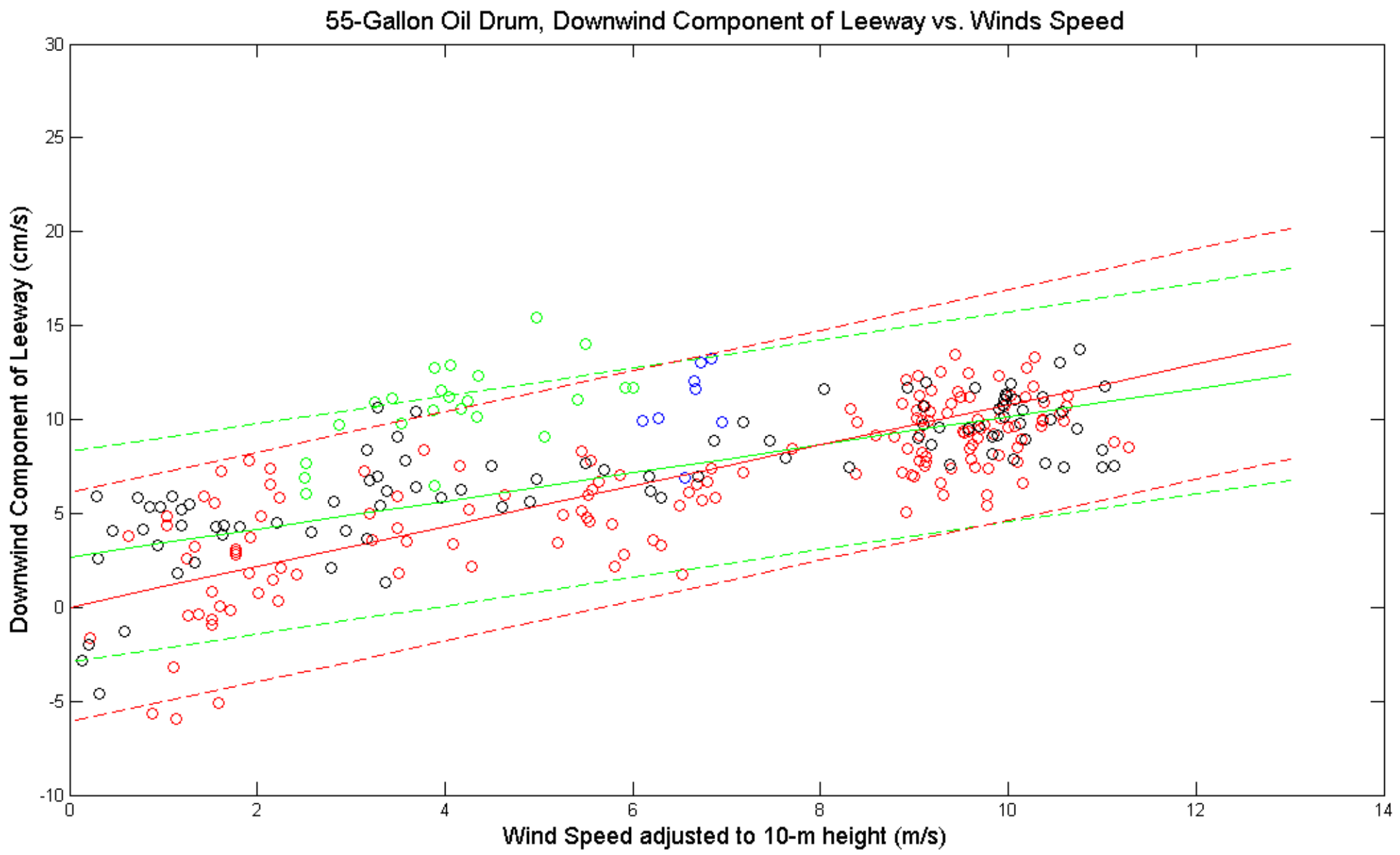

Figure 9. Downwind Component of Leeway (cm $s^{-1}$) *v* wind speed adjusted to 10-m height for the 220-l (55-gallon) oil drum. Unconstrained linear regression (solid) and 95% confidence levels (dash) are plotted in green. Constrained linear regression is plotted in red. Run 1 data are plotted in blue, Run 2 in red, Run 3 in green and Run 4 in black.

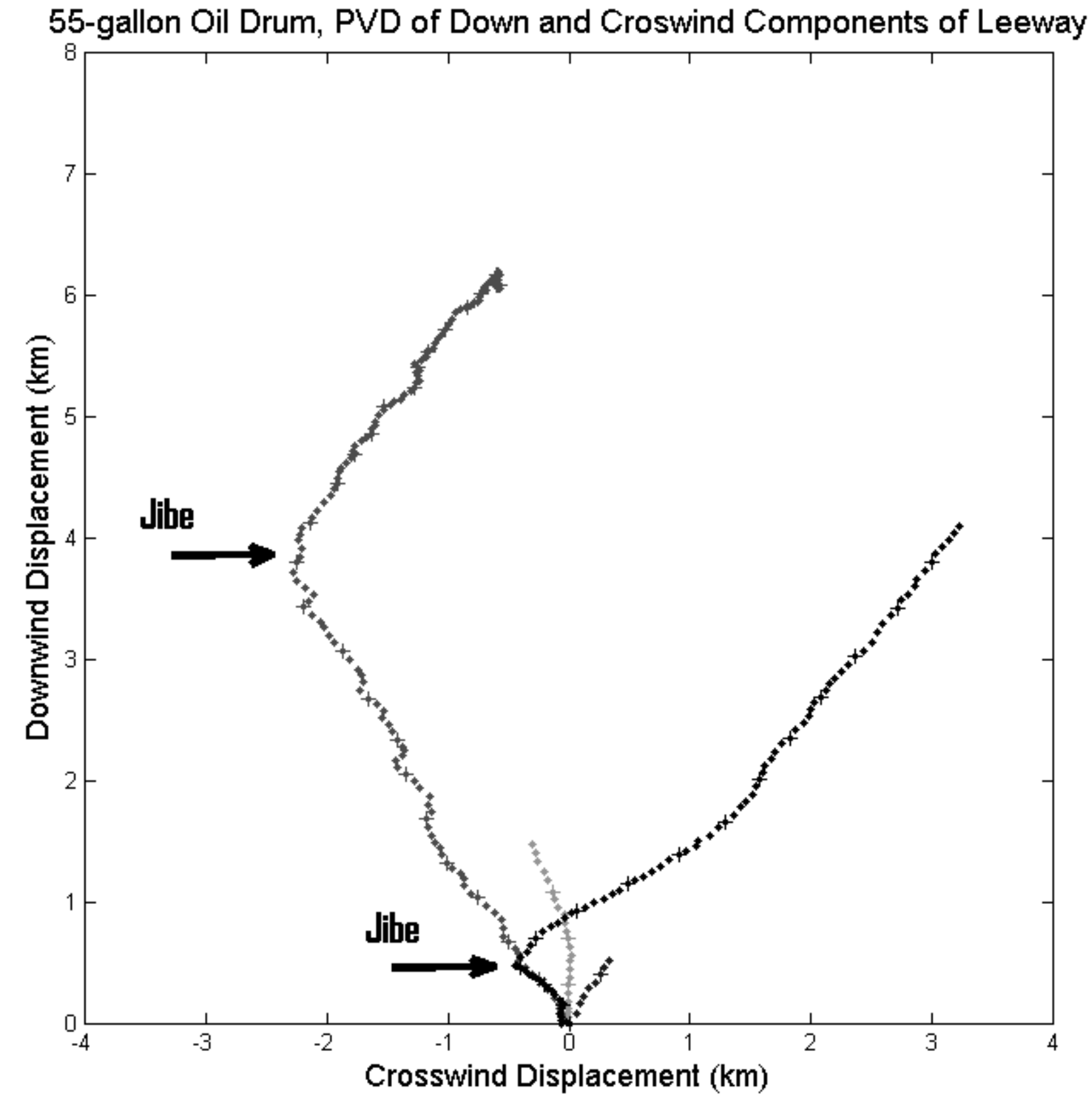

Figure 10. Progressive Vector Diagram (PVD) of the downwind and crosswind components of leeway displacement, for the 220-l (55-gallon) drum. Downwind direction is up, positive crosswind to the right, displacement is in kilometers. Hourly intervals are indicated by a hatched point every 6th point. Horizontal black arrows indicate the two major shifts in the sign (jibes) of the crosswind component of leeway that were found in the four runs. The presence of jibes means that the analysis of cross-wind leeway (CWL) must be performed after data have been separated into right and left-drifting events (see Table 3).

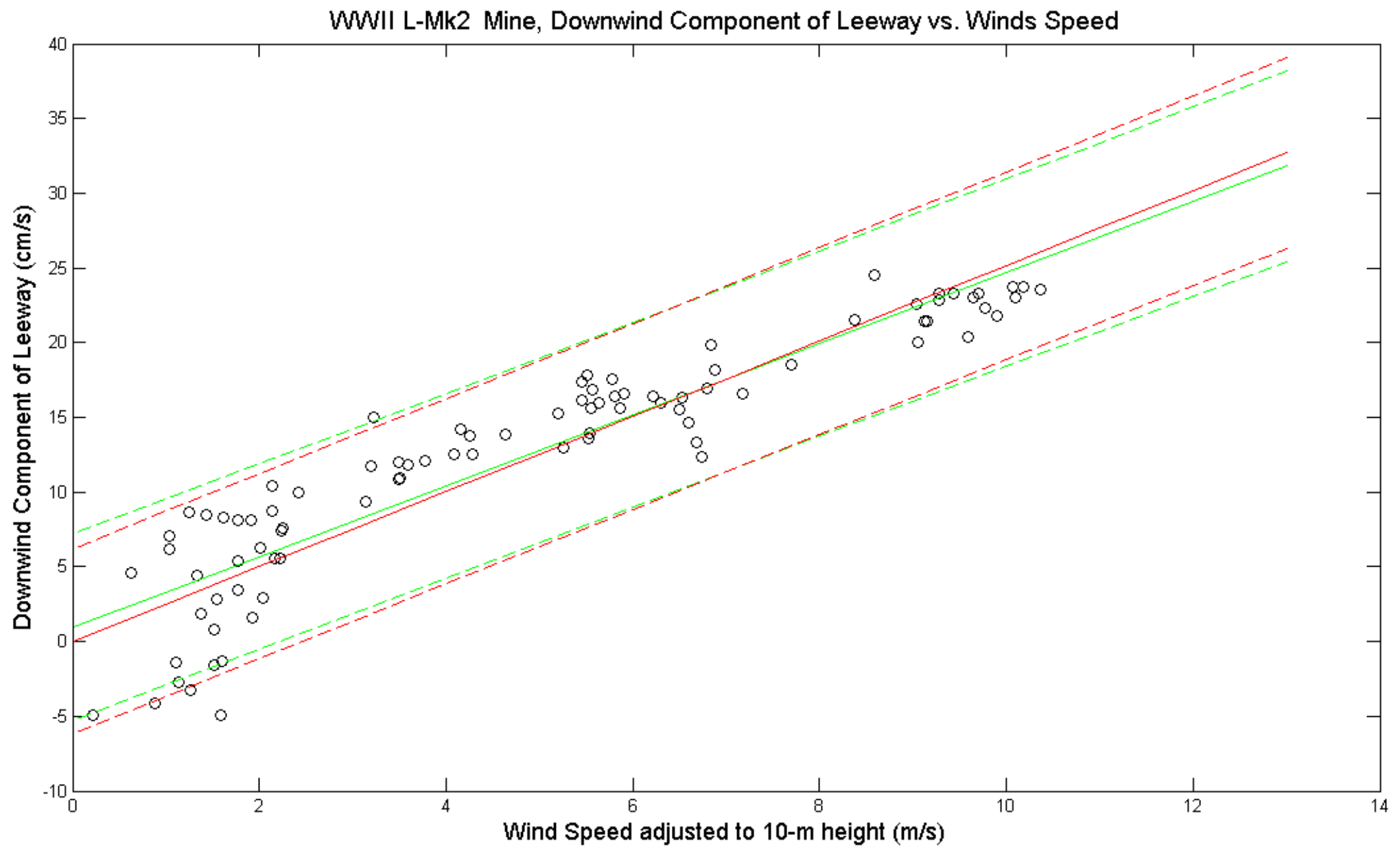

Figure 11. Downwind component of leeway (cm s$^{-1}$) *v* wind speed adjusted to 10 m height for the WWII L-MK2 mine. Unconstrained linear regression mean (solid) and 95% confidence levels (dashed) are plotted in green. Constrained linear regression is plotted in red.

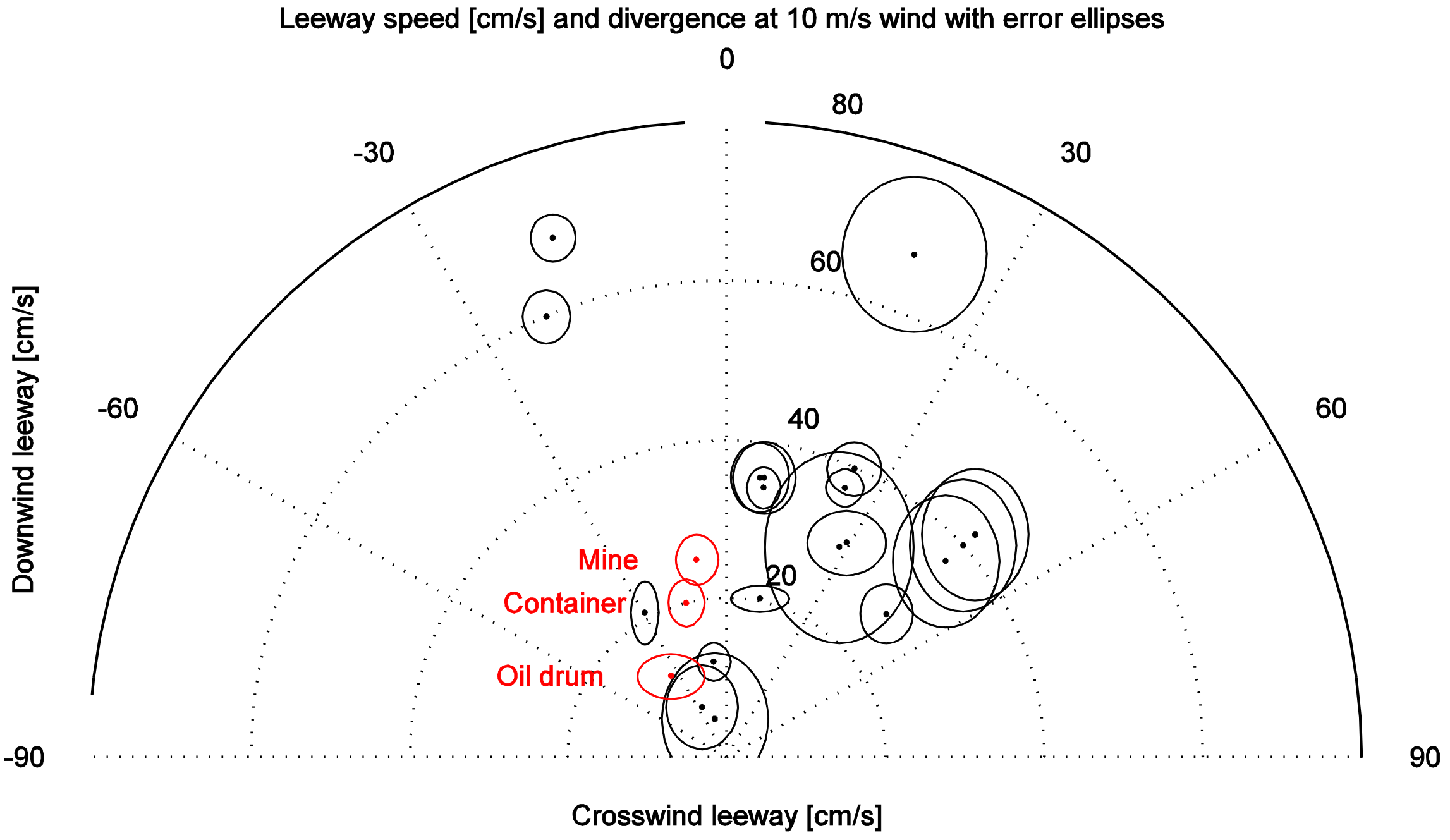

Figure 12. Distribution of the three objects studied in the two-dimensional leeway-space (red) at 10 m $s^{-1}$ wind speed. The object categories from Figure 1 are reproduced in black for comparison. The distance from the origin indicates leeway speed [cm $s^{-1}$] while the angle relative to the vertical indicates the object's divergence from the wind direction (wind blowing upwards along second axis). The ellipses show crosswind and downwind error. Categories are placed selectively in the left and right quadrant for readability.

# Tables

| Object | Leeway Speed | | | | Downwind leeway (DWL) | | | | Crosswind leeway (CWL) | | |
|---|---|---|---|---|---|---|---|---|---|---|---|
| | Slope (%) | Offset (cm $s^{-1}$) | $S_{yx}$ (cm $s^{-1}$) | $r^2$ | Slope (%) | Offset (cm $s^{-1}$) | $S_{yx}$ (cm $s^{-1}$) | $r^2$ | Slope (%) | Offset (cm $s^{-1}$) | $S_{yx}$ (cm $s^{-1}$) |
| Container | 1.4 | 4.8 | 3.0 | 0.92 | 1.8 | 1.4 | 3.0 | 0.84 | 0.27 | 2.4 | 2.3 |
| Drum | 0.76 | 5.3 | 2.9 | 0.44 | 0.75 | 2.7 | 2.8 | 0.45 | 0.47 | 2.3 | 4.2 |
| Mine | 2.0 | 3.9 | 1.9 | 0.92 | 2.4 | 0.9 | 3.1 | 0.85 | 0.18 | 2.0 | 2.6 |

Table 1. Unconstrained Linear Regression Parameters for the three objects studied. The wind range covered is found in Table 2.

| Object | 10-m wind speed range (m $s^{-1}$) | No of 10 min samples | Leeway speed | | | DWL | | | CWL | |
|---|---|---|---|---|---|---|---|---|---|---|
| | | | Slope (%) | $S_{yx}$ (cm $s^{-1}$) | $r^2$ | Slope (%) | $S_{yx}$ (cm $s^{-1}$) | $r^2$ | Slope (%) | $S_{yx}$ (cm $s^{-1}$) |
| Container | 0-11 | 95 | 2.0 | 3.0 | 0.71 | 2.0 | 3.1 | 0.83 | 0.02 | 2.7 |
| Drum | 0-11 | 278 | 1.4 | 3.9 | 0.12 | 1.1 | 3.1 | 0.34 | 0.76 | 4.3 |
| Mine | 0-10 | 89 | 2.6 | 2.8 | 0.80 | 2.5 | 3.1 | 0.84 | 0.13 | 2.8 |

Table 2. Linear regression parameters constrained through zero.

| | Positive CWL ($N$=164 10 min samples) | | | Negative CWL ($N$=114 10 min samples) | | |
|---|---|---|---|---|---|---|
| | Slope (%) | Offset (cm $s^{-1}$) | $S_{yx}$ (cm $s^{-1}$) | Slope (%) | Offset (cm $s^{-1}$) | $S_{yx}$ (cm $s^{-1}$) |
| Oil Drum Unconstrained | 0.49 | 2.9 | 3.9 | -0.45 | -1.5 | 4.6 |
| Oil Drum Constrained | 0.86 | N/A | 4.1 | -0.62 | N/A | 4.6 |

Table 3. Separate regressions of right-drifting (positive CWL) and left-drifting (negative CWL) 10-minute events for the oil drum, constrained and unconstrained. The total average of left and right-drifting events is found in Table 1.